\documentclass[useAMS]{mn2e}
\usepackage{graphicx}
\usepackage{epsfig}
\usepackage{amssymb}
\usepackage{lscape}
\usepackage{ulem}
\usepackage{txfonts}

\def\sigh2{$\Sigma_{\rm H_2}$}

\def\kms{km~s$^{-1}$}

\def\c2s{C\,{\sc ii}$^{\star}$}

\title[Post-merger quenching] {Galaxy mergers can rapidly shut down star formation}

\author[Ellison et al.] {Sara L. Ellison$^1$, Scott Wilkinson$^1$, Joanna Woo$^2$, Ho-Hin Leung$^3$, Vivienne Wild$^3$, 
  \newauthor Robert W. Bickley$^1$, David R. Patton$^4$, Salvatore Quai$^1$, Stephen Gwyn$^5$ \\
$^1$ Department of Physics \& Astronomy, University of Victoria, Finnerty Road, Victoria, British Columbia, 
  V8P 1A1, Canada \\
$^2$ Department of Physics, Simon Fraser University, 8888 University Drive, Burnaby BC V5A 1S6, Canada\\
$^3$ School of Physics and Astronomy, University of St Andrews, North Haugh, St Andrews, KY16 9SS, U.K.\\
$^4$ Department of Physics and Astronomy, Trent University, 1600 West Bank Drive, Peterborough, ON K9L 0G2, Canada\\
$^5$ NRC Herzberg Astronomy and Astrophysics, 5071 West Saanich Road, Victoria, BC V9E 2E7, Canada
}

\begin{document}

\maketitle

\begin{abstract}

Galaxy mergers trigger both star formation and accretion onto the central supermassive black hole.  As a result of subsequent energetic feedback processes, it has long been proposed that star formation may be promptly extinguished in galaxy merger remnants.  However, this prediction of widespread, rapid quenching in late stage mergers has been recently called into question with modern simulations and has never been tested observationally.  Here we perform the first empirical assessment of the long-predicted end phase in the merger sequence.  Based on a sample of $\sim$ 500 post-mergers identified from the Ultraviolet Near Infrared Optical Northern Survey (UNIONS), we show that the frequency of post-merger galaxies that have rapidly shutdown their star formation following a previous starburst is 30-60 times higher than expected from a control sample of non-merging galaxies.  No such excess is found in a sample of close galaxy pairs, demonstrating that mergers can indeed lead to a rapid halt to star formation, but that this process only manifests after coalescence.   
  
\end{abstract}


\section{Introduction}\label{intro_sec}

The motions of galaxies through space can sometimes lead to gravitational interactions which result in a merger (e.g. Hopkins et al. 2008; Lotz et al. 2011).  It has long been known that the interaction process can have profound effects on galactic structure (e.g. Toomre \& Toomre 1972; Patton et al. 2016), star formation (e.g. Ellison et al. 2008; Scudder et al. 2012) and fuelling of the central supermassive black hole (Ellison et al. 2011; Satyapal et al. 2014).  Following intense bursts of merger-triggered star formation and black hole accretion, it has been theorized that feedback/gas removal resulting from these energetic processes may lead to the subsequent rapid shutdown of further star formation in the post-merger remnant (Springel et al. 2005; Hopkins et al. 2008).  However, this prediction of widespread, rapid cessation of star formation in late stage mergers has been recently called into question with modern simulations (Zheng et al. 2020; Quai et al. 2021) and has never been tested observationally.

The category of galaxies known as `post-starbursts’ is characterized by having recently experienced an episode of intense star formation that was rapidly truncated (e.g. Dressler \& Gunn 1983; Couch \& Sharples 1987).  Post-starburst galaxies therefore represent an excellent tool for testing whether galaxy mergers can lead to a prompt quenching of star formation.  The nature of post-starbursts, and their connection to galaxy mergers, has been extensively studied (e.g. Zabludoff et al. 1996; Goto 2005; Pawlik et al. 2016; Wilkinson et al. 2022).  These studies have demonstrated that a significant fraction of post-starbursts exhibit disturbed morphologies, with a merger fraction several times in excess of the expectations from a control sample (see Wilkinson et al. 2022 for a comprehensive breakdown of post-starburst merger fractions using a variety of selection metrics).  The high fraction (and excess) of merger activity amongst the post-starburst population demonstrates that merging is one potential avenue to quenching.

All previous studies of the post-starburst-merger connection have started with a post-starburst sample and examined images (visually or with automated tools) to assess the merger fraction amongst this recently quenched population.  However, the lack of a large, homogeneously selected sample of fully coalesced post-mergers has precluded inverting the question and asking `what fraction of mergers host a post-starburst?'  It is this latter question that we aim to answer in the current work, hence testing whether mergers \textit{systematically} lead to the quenching of star formation.

There are three principal challenges in the compilation of a large, representative post-merger sample:  1) the need for a large survey to identify the inherently rare recent post-merger phase; 2) the requirement of high-quality imaging in order to detect the (often faint) tell-tale disturbance from a gravitational interaction and 3) the need for automated classification of post-mergers to handle these large datasets.

The Canada France Imaging Survey (CFIS), which is part of the Ultraviolet Near Infrared Optical Northern Survey (UNIONS), addresses the first two of these challenges, by providing deep $r$-band imaging that will eventually extend over $\sim$5000 degrees of the sky. The third challenge can be addressed with the development of machine learning methods, in particular through the application of convolutional neural networks (CNN), to identify galaxy mergers (Ackermann et al. 2018; Bottrell et al. 2019; Ferreira et al. 2020).  Using a bespoke CNN that is tailored to the specifics of CFIS (Bickley et al. 2021), with subsequent visual confirmation, we have identified a highly pure sample of 699 post-mergers in the CFIS 2nd data release (Bickley et al. 2022).  By combining the CFIS post-merger catalog with extant post-starburst classifications available in the literature, in the work presented here we make the first statistical assessment of the frequency of rapid quenching in recently coalesced mergers.

\section{Methods}

\subsection{Selection of the parent galaxy sample}\label{parent_sec}

In order to assemble a parent sample for which we have access to both the high quality optical ($r$-band) imaging required for merger identification, as well as the spectroscopy required to identify the post-starburst population and measurements of galactic properties such as stellar mass and star formation rate, we cross-match the Sloan Digital Sky Survey (SDSS) data release 7 (DR7) main galaxy sample with objects identified in the CFIS data release 2 (DR2).  The stellar masses and SFRs used in this work are taken from the MPA/JHU catalog (Kauffmann et al. 2003a; Brinchmann et al. 2004). Approximately 168,000 galaxies are successfully matched between CFIS and SDSS within a positional tolerance of 3 arc seconds.  Following previous work (e.g. Goto 2007; Pawlik et al. 2018; Wilkinson et al. 2022), in order to ensure adequate signal-to-noise (S/N) to permit a robust post-starburst classification, we further require a g-band S/N $>$ 8 in the SDSS spectrum, which reduces the sample to 99,703 galaxies.  Finally, since the CNN (see below) used to identify post-merger galaxies in CFIS (Bickley et al. 2021) was trained using simulated images of galaxies with stellar masses M$_{\star} > 10^{10}$ M$_{\odot}$, we likewise apply this mass cut to the sample.  The final parent sample used in this work therefore consists of 81,899 galaxies which all have high quality CFIS $r$-band imaging, high S/N SDSS spectroscopy and are massive enough for a robust merger classification.

\subsection{CNN selection of post-merger galaxies}

The CNN used to identify galaxy post-mergers is described in detail in our previous work (Bickley et al. 2021).  In brief, we used the IllustrisTNG (TNG100-1) cosmological hydrodynamical simulation (e.g. Pillepich et al. 2018) to identify galaxies that have merged between two consecutive time steps (typically sampling $\sim$150 Myrs).  The selection of post-mergers follows the steps described in Hani et al. (2020).  Specifically, the merger must have occurred between two galaxies whose stellar masses were within a factor of 10 of one another and since a redshift of one.  Given the stellar particle mass of the TNG100-1 simulation, galaxies are well resolved if they have a stellar mass above $\sim$10$^9$ solar masses.  We therefore impose a minimum stellar mass of 10$^{10}$ solar masses on the post-merger sample, in order to have a complete sample of mergers with mass ratios of at least 1:10.  A sample of non-mergers (no major merger within at least 2 Gyr) that is matched to the post-merger sample in redshift, stellar mass and environment is also generated.  Mock CFIS $r$-band images are produced from the simulation stellar mass maps in order to replicate the realistic seeing and noise conditions of the survey, an essential ingredient in successful CNN training (Bottrell et al. 2019).  The trained CNN yields an overall classification accuracy of 88 percent (Bickley et al. 2021).

Application of the CNN to the complete CFIS $r$-band coverage (without our additional cuts in stellar mass or spectral S/N) yields 2000 galaxies with a post-merger probability of at least 0.75 (Bickley et al. 2022).  However, due to the intrinsic rarity of post-mergers, Bayes theorem predicts that even a classifier with accuracy as high as 90 percent will still only be 6 percent pure at the default decision threshold of 0.5 (Bickley et al. 2021; Bottrell et al. 2022).  We have therefore performed an additional visual check on all (2000) galaxy images with a CNN prediction $>$0.75, resulting in a highly pure sample of 699 post-mergers (Bickley et al. 2022).  The post-merger sample is characterized by single galaxies (with no companion or second nucleus) that exhibit a disturbed morphology, and thus represents the post-coalescence phase of the interaction. Of this visually cleaned, gold-standard CFIS post-merger sample, 508 galaxies additionally fulfill our requirements of SDSS $g$-band S/N$>$8 and log M$_{\star}$ $>$ 10 M$_{\odot}$.

\subsection{Selection of pre-coalescence galaxy pairs from SDSS spectroscopy}

In order to study the pre-coalescence phase of the merger sequence, we additionally identify a sample of (spectroscopic) galaxy pairs.  The pair sample is also selected from the SDSS DR7 (Patton et al. 2016), with the requirements that the stellar mass ratio of the two galaxies is within a factor of 10, the projected separation is less than 80 kpc and the line of sight rest frame velocity difference is less than 300 \kms.  For fair comparison with the post-merger sample, we again require that the stellar mass is above 10$^{10}$ solar masses and that the SDSS $g$-band S/N$>$8.  15,979 pairs are thus selected.  Since the identification of galaxy pairs relies only on the SDSS and is independent of CFIS imaging, no overlap with the latter survey is required.

\subsection{Identification of non-merger control samples}

A control sample of non-merging galaxies is required in order to assess whether post-mergers occur more frequently than expected in the general galaxy population.  Controls are drawn from the parent sample described in Section \ref{parent_sec}.  However, in order to remove possible mergers from the parent sample we first reject any galaxy with a CNN merger prediction $>$0.5.  Based on the naturally occurring frequency of post-mergers in the universe and the CNN’s classification performance, we expect the frequency of `contaminating’ post-mergers in this pool of potential control non-merging galaxies to be $<$0.1 per cent.  Of the initial $\sim$ 82,000 galaxies in the parent sample, 78,295 remain as non-mergers.

For each of the 508 post-mergers in turn, we select the closest match in stellar mass from the non-merger parent sample, without replacement.  In order to build a statistically robust comparison sample, this process is repeated 50 times yielding a control sample of 25,400 galaxies.  A Kolmogorov-Smirnov (KS) test confirms that the distribution of stellar masses in the post-merger and control samples is indistinguishable.

A mass-matched control sample is also assembled for the pairs.  The parent sample from which these controls is drawn includes all SDSS galaxies whose closest spectroscopic companion is $>$ 100 kpc away in projected separation.  227,147 galaxies without a close companion form the pool from which controls are drawn.  We follow the same methodology as matching the controls for the post-merger sample; the closest mass-matched galaxy is identified for each pair without replacement.  Due to the much larger sample of pairs compared with post-mergers, the number of controls per pair galaxy is set to be ten, resulting in a total of 159,790 control galaxies for the pair sample. 

\subsection{Quantitative morphologies}

Beyond the simple CNN classification of a galaxy as a post-merger, it is useful to have a further continuous metric of the extent of the tidal disturbance for each galaxy. We measure quantitative morphologies for the entire CFIS sample using the publicly available Statmorph code (Rodriguez-Gomez et al. 2019).  The preparation of the images, object definition, masking procedures and all relevant details pertaining to the morphological parameterization are discussed in detail in Wilkinson et al. (2022).  In the work presented here, we make use of the shape asymmetry parameter (Pawlik et al. 2016).  In contrast to the traditional asymmetry metric (which sums the residual flux in a 180-degree rotated image of the galaxy subtracted from the original; Conselice et al. 2000), shape asymmetry uses a binary mask, in order to enhance the impact of faint tidal features.  Although shape asymmetry (as well as other metrics such as traditional asymmetry and Gini-M20) can be used independently to identify mergers, CNN have been found to perform at a superior level when identifying post-merger samples (Bickley et al. 2021; Wilkinson et al. 2022).  In the work presented here, we therefore use shape asymmetry only as a metric of the extent of structural disturbance, rather than for identifying the initial sample of post-mergers.

\subsection{Post-starburst classifications}

Two complementary methods for post-starburst selection are used in the work presented here.  Although both aim to identify, via their strong Balmer absorption features, galaxies that have recently and abruptly truncated their star formation following a burst, the two methods contrast in their permission of emission lines from non-star-forming processes.

The first post-starburst selection used in the current work adopts the traditional approach of combining strong Balmer absorption with the absence of emission lines, a method that yields a classic `E+A’ sample of post-starbursts (e.g. Zabludoff et al. 1996; Goto 2005).  Specifically, we use the previously published sample of E+A post-starbursts identified in the SDSS DR7 (Goto 2007) which requires the following equivalent width (EW) thresholds (where positive values indicate absorption and negative values indicate emission): EW(H$\delta$)$>$ 5.0 \AA, EW([OII])$>-$2.5 \AA\ and EW(H$\alpha$)$>-$3.0 Å.  The E+A sample therefore contains galaxies that have experienced a strong recent burst (as evidenced by the Balmer absorption indicative of A and F type stars) but no on-going star formation (absence of O and B stars as evidenced by the lack of recombination lines in emission).

The drawback of the E+A selection method is that it removes galaxies whose emission lines may originate from a non-star-forming origin, such as from shocks or an active galactic nucleus (AGN), as well as galaxies that haven’t completely shut-down their star formation.  Several alternative selection methods have therefore been proposed in order to identify bona fide post-starbursts with emission lines (e.g. Mendel et al. 2013; Yesuf et al. 2014; Alatalo et al. 2016).  In the work presented here, we use a principal component analysis (PCA; Wild et al. 2007) which selects galaxies with a relative excess of Balmer absorption given the age of their stellar population.  Specifically, we use the PCA catalog published for SDSS DR7 (Wild et al. 2007).  In addition to cuts in PCA space, we additionally apply mass dependent thresholds in Balmer decrement designed to remove dusty galaxies that may not be true post-starbursts.  The complete set of spectral requirements for the PCA post-starburst sample is described in detail in Wilkinson et al. (2022).

\section{Results}

\begin{figure}
	\includegraphics[width=8cm]{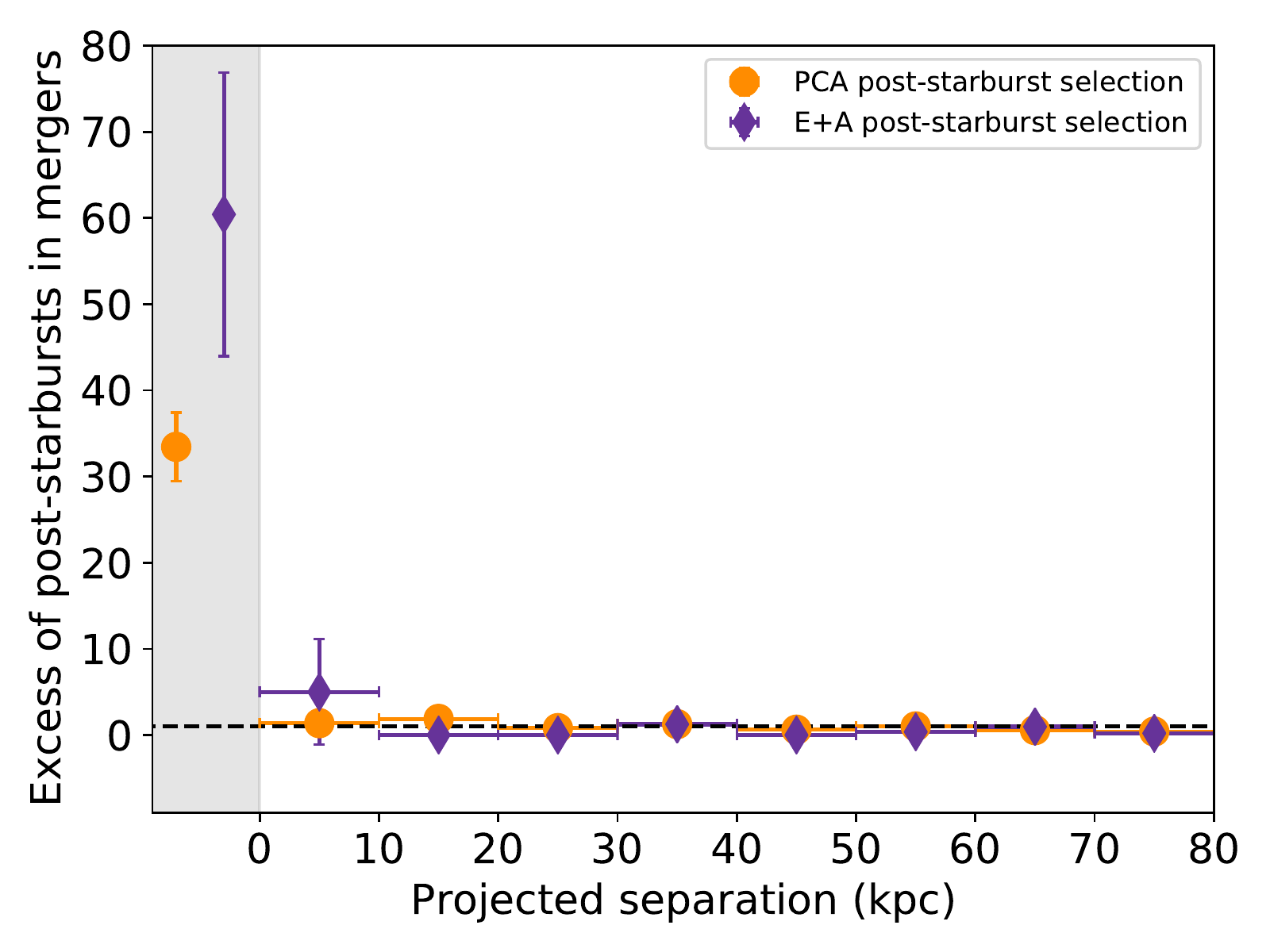}
        \caption{The excess of post-starbursts in mergers, as defined by the fraction of post-starbursts in mergers compared with their mass-matched controls.  The data points in the grey shaded box indicate the post-starburst excess in the post-merger sample.  Other data points are for close galaxy pairs, plotted as a function of the projected separation to their companion.  The statistics are computed for two methods of post-starburst selection, the traditional E+A method (purple diamonds) and a PCA approach (orange circles).  The horizontal dashed line shows a value of 1.0, i.e. an identical fraction of post-starbursts in the mergers and their controls. X-axis and y-axis error bars show bin widths and binomial uncertainties, respectively.}
    \label{psb_excess}
\end{figure}

In order to quantify the fraction of galaxies that have recently and rapidly shut down their star formation in the post-merger and control samples, we compute the fraction of post-starbursts in each, using both the E+A and the PCA selection methods.  We find that 6$\pm$1 percent of post-mergers and 0.10$\pm$0.02 percent of controls are classified as post-starbursts with the E+A technique.  Using the PCA post-starburst classification these fractions are increased to 20$\pm$2 percent and 0.60$\pm$0.05 percent for post-mergers and controls respectively, where the higher absolute fraction is expected for the more inclusive (of emission lines) selection of the PCA approach.  \textit{Thus, we find that rapid truncation of star formation following a burst is 30 - 60 times more common in post-mergers than expected in a control sample}, depending on the technique used to identify the post-starburst.  This excess of post-starbursts in post-mergers exists not only for the sample as a whole, but also at fixed specific SFR.  Therefore, although the fraction of post-starbursts in post-mergers is not dominant in an absolute sense, the merger process is clearly conducive to completely shutting down star formation (at least temporarily – later re-ignition of star formation may still be possible).  Moreover, we note that the quoted percentages of post-starbursts in post-mergers are likely to be an under-estimate, because 1) the post-starburst selection methods are most sensitive to relatively high burst mass fractions (Wild et al. 2007) and 2) the timescale for merger feature visibility is often shorter than the post-starburst ages (Pawlik et al. 2018; indeed, merger features may fade even before the post-starburst signature is visible).   

In order to claim that the merger is leading to quenching in the post-coalescence regime, it is necessary to confirm that the pre-coalescence status of the galaxy was one of active star formation.  The alternative scenario is that the interacting galaxies were already quiescent, but the merger led to a brief period of rejuvenated star formation before returning to its formerly quenched state.  Using the spectral fitting techniques described in Wild et al. (2020) that have been optimized for post-starbursts, along with improvements for two-component modelling as described in Leung et al. (in prep), we have derived the (in-fibre) SFRs of post-merger galaxies in the pre-burst epoch (1-2 Gyrs ago).  The vast majority (98 percent) of post-merger post-starbursts in our sample have pre-burst fibre SFRs $>$ 1 M$_{\odot}$/yr (a value consistent with main sequence star formation in the mass range of our sample), confirming that we are witnessing a true transition from star-forming to quiescence as a result of the merger.

Many of the transformations in galaxy behaviour (such as increased SFR and enhanced black hole accretion) are already detectable in pre-coalescence galaxy pairs (Ellison et al. 2008, 2011; Scudder et al. 2012).  In Figure \ref{psb_excess}, we explore whether the post-starburst excess that we have measured in post-merger galaxies is also present in our sample of close galaxy pairs.  The excess frequency of post-starbursts (defined as the fraction of post-starbursts in mergers divided by the fraction of post-starbursts in the control sample) as measured by both the E+A and the PCA selection methods is shown for the pairs and the post-merger samples.   Although there is a hint of an excess of post-starbursts in the closest separation pairs (0-10 kpc), this is not statistically significant.  Figure \ref{psb_excess} therefore demonstrates that the shut-down of star formation occurs primarily in the post-merger regime.

\begin{figure}
	\includegraphics[width=8cm]{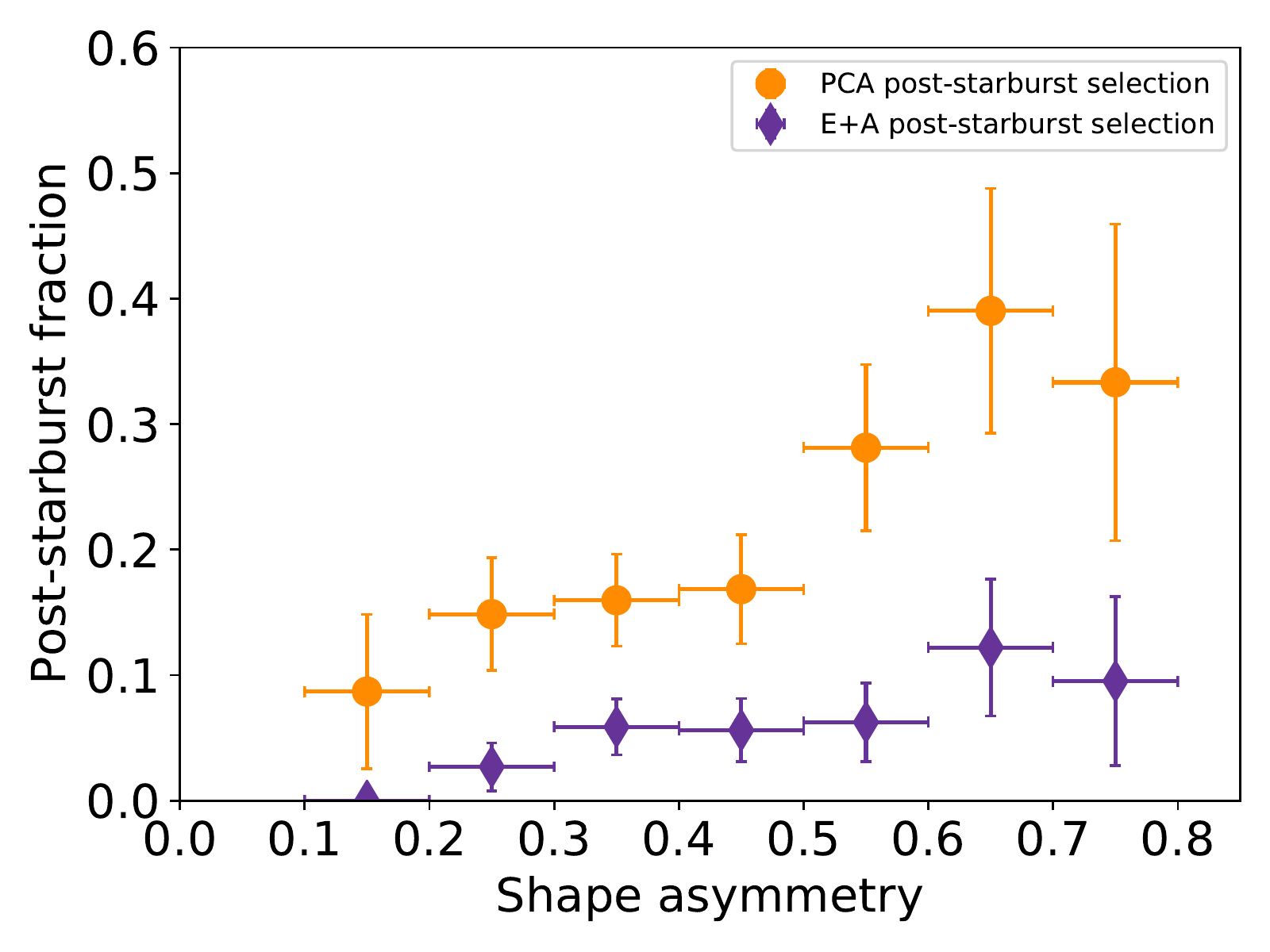}
        \caption{The post-starburst fraction of post-mergers as a function of the galactic shape asymmetry.  High shape asymmetries indicate more disturbed morphologies.  Post-mergers that are more highly disturbed have 4-5 times higher fractions of both PCA-selected (orange circles) and E+A-selected (purple diamonds) post-starbursts, compared with low asymmetry post-mergers. X-axis and y-axis error bars show bin widths and binomial uncertainties, respectively. }
    \label{psb_asym}
\end{figure}

Simulations of galaxy interactions (in either idealized or cosmological settings) generally agree that whilst mergers can lead to a post-starburst phase, this outcome requires particularly favourable conditions, such as prograde orbits, similar stellar masses, high gas fractions and disk-dominated morphologies (Wild et al. 2009; Snyder et al. 2011; Zheng et al. 2020).  Although we are not able to reconstruct the pre-merger conditions for our observed sample, many of the properties that are predicted to be conducive to the generation of a post-starburst tend to also produce the most dramatically disturbed post-merger remnants (Lotz et al. 2010a,b; Pawlik et al. 2018; Nevin et al. 2019).   In Figure \ref{psb_asym}, we therefore investigate the dependence of the post-starburst fraction on the extent of the galaxy’s morphological disturbance as quantified by its shape asymmetry for the 508 post-mergers in our CFIS sample.   Figure \ref{psb_asym} shows a correlation between the asymmetry of the post-merger galaxies and the frequency of post-starbursts therein.  Specifically, at the highest shape asymmetries, 40 percent of post-mergers are classified as post-starbursts using the PCA selection method, which is double the frequency of the sample taken as a whole and four times the enhancement seen for the lowest asymmetries.  The correlation between post-starburst fraction and asymmetry is also seen in the E+A selection, increasing from 2 percent at low asymmetries to 10 percent in the most disturbed galaxies. The results in Figure \ref{psb_asym} are thus consistent with a scenario in which the most morphologically disruptive merger events are the ones leading preferentially to a rapid halt to star formation.  This interpretation is underscored by the fact that most mergers only exhibit enhanced shape asymmetries for $\sim$ 0.5 Gyrs (Pawlik et al. 2018; although this timescale will depend on image depth) whereas most PCA-selected post-starbursts are expected to have ages between  $\sim$ 0.5 and 1 Gyr.  Therefore, for most mergers, the timescales for measuring high shape asymmetry are incompatible with those for detecting PCA post-starbursts. A correlation between shape asymmetry and the frequency of post-starbursts is thus expected when the most favourable conditions for shutting down star formation are the same as those that extend the visibility timescale of the morphological disturbance.

\section{Discussion}

Although our study has demonstrated, for the first time, a strong statistical excess of post-starbursts in post-merger galaxies, the internal physical mechanism that leads to the shutdown of star formation is not addressed by our analysis.  Although gas exhaustion (consumption by the starburst) or removal (by energetic feedback) are obvious culprits, several studies of post-starbursts have demonstrated that ample reservoirs of molecular gas still remain in many low redshift post-starbursts (French et al. 2015; Rowlands et al. 2015; Yesuf \& Ho 2020).  However, the lack of molecular gas in its densest phases (French et al. 2018) and enhanced turbulence (Smercina et al. 2022) indicate that some mechanism is preventing the gas from achieving the conditions necessary to form stars.   Feedback from AGN is often implicated as one such mechanism in the general galaxy population (Terrazas et al. 2017; Martin-Navarro et al 2018; Bluck et al. 2020).  AGN feedback may contribute in multiple ways, ranging from a small scale impact on circumnuclear gas (Izumi et al. 2020; Ellison et al. 2021; Ramos-Almeida et al. 2022) to a galaxy-wide effect based on the integrated energy deposited into the halo (Terrazas et al 2020; Zinger et al. 2020; Piotrowska et al. 2022), the so-called `preventative’ feedback mode.  Indeed, some observational studies have identified significant AGN fractions in post-starburst samples (Goto 2006; Yesuf et al. 2014; Pawlik et al. 2018).  However,  we find that although 21 percent of the PCA-defined post-starburst post-merger sample are classified as AGN based on their optical emission line ratios (Kauffmann et al. 2003b), 27 percent of the non-post-starburst post-mergers are also AGN.  The prevalence of mid-IR identified AGN (using a threshold of W1$-$W2$>$0.5 from the Wide-field Infrared Survey Explorer) are also similar between post-starburst and non-post-starburst post-mergers.  We therefore do not find that AGN are preferentially linked with those post-mergers which are currently in a post-starburst phase.  We suggest that the high fraction of post-starbursts that have previously been found to host AGN is a combined result of 1) a high fraction of post-starbursts being morphologically disturbed (i.e. likely ongoing mergers; Zabludoff et al. 1996; Goto 2005; Wilkinson et al. 2022) and 2) the enhanced frequency of AGN in mergers (e.g. Ellison et al. 2011, 2019; Satyapal et al. 2014).  The enhanced turbulence seen in molecular gas of post-starbursts (Smercina et al. 2022) and the subsequent inability of the remaining molecular gas to form stars may alternatively be linked directly to the dynamical impact of the merger.


\section*{Data Availability}

A subset of the raw CFIS imaging data are publicly available via the Canadian Astronomical Data Center at http://www.cadc-ccda.hia-iha.nrc-cnrc.gc.ca/en/megapipe/. The remaining raw data and all processed data are available to members of the Canadian and French communities via reasonable requests to the principal investigators of the Canada–France Imaging Survey, Alan McConnachie and Jean-Charles Cuillandre. All data will be publicly available to the international community at the end of the proprietary period, scheduled for 2023.  The catalog of post-mergers from which we have constructed our sample is publicly available as an online supplement to its original journal publication13.  Stellar masses, SFRs and measurements of line equivalent widths are all publicly from the MPA/JHU SDSS DR7 catalogs: https://wwwmpa.mpa-garching.mpg.de/SDSS/DR7/.  Principal components of SDSS DR7 galaxies are available here: http://www-star.st-and.ac.uk/~vw8/downloads/DR7PCA.html.  The sample of E+A galaxies in the SDSS DR7 is available here: http://www.phys.nthu.edu.tw/~tomo/cv/index.html.

\end{document}